\documentclass[12pt]{iopart2}
\usepackage{amsmath,amssymb}

\usepackage{bm,mathrsfs,color}
\usepackage{graphicx}
\usepackage{braket}

\newcommand{\bv}{\bm{v}}

\newcommand{\bk}{\bm{k}}

\newcommand{\kp}{\bm{k}\cdot \bm{p}}

\newcommand{\ve}{\varepsilon}

\begin{document}

%\title[]{Longitudinal and transverse magnetoresistance in SrTiO$_3$}
\letter{
%\begin{center}{\normalsize \underline{{\tt ver. 0.6} {\normalsize {\tt (\today)}}}} \vspace{5mm}\end{center}
Longitudinal and transverse magnetoresistance of SrTiO$_3$ with a single closed Fermi surface}

\author{Yudai Awashima$^1$ \& Yuki Fuseya$^{1, 2}$}

\address{
$^1$ Department of Engineering Science, University of Electro-Communications, Chofu, Tokyo 182-8585, Japan}
\address{$^2$ Institute for Advanced Science, University of Electro-Communications, Chofu, Tokyo 182-8585, Japan}
\ead{awashima@kookai.pc.uec.ac.jp}
\vspace{10pt}
%\begin{indented}
%\item[]August 2017
%\end{indented}

\begin{abstract}
The magnetoresistance (MR) of SrTiO$_3$ is theoretically investigated based on the Boltzmann equation by considering its detailed band structure. The formula for MR proposed by Mackey and Sybert is extended to be applicable to a system with an arbitrarily shaped Fermi surface. It is shown that the angular dependence of the diagonal component of the mass tensor causes transverse MR, whereas that of the off-diagonal component causes longitudinal MR with only a single closed Fermi surface, which overturns the textbook understanding of MR. The calculated MR (300\% at 10 T) quantitatively agrees with the experimental results for SrTiO$_3$ including the behavior of the linear MR. The negative Gaussian curvature of the Fermi surface of SrTiO$_3$ and its resulting negative longitudinal and transverse MR are also discussed.
\end{abstract}

%\maketitle

\section{Introduction}
Magnetoresistance (MR) has been the subject of many studies in solid state physics since its discovery by Thomson \cite{Thomson1857}. 
Despite the long history of studies on MR \cite{Pippard_book,ZZhu2018}, there are still some questions regarding its fundamental mechanisms. According to standard textbooks on solid state physics \cite{Ziman_book,Grosso_Parravicini_book}, the resistivity perpendicular to an applied magnetic field (transverse MR, $\rho_{xx}$) is independent of the magnetic field for a spherical one-band model with a single relaxation time. However, experimental results have shown nearly all metals with a single closed Fermi surface exhibit transverse MR. 

%============================================================
\begin{figure}
	\begin{center}
		\includegraphics[width=8cm]{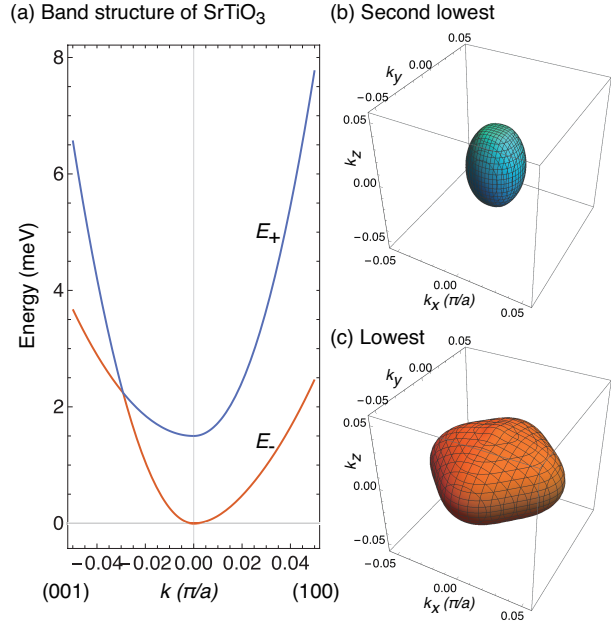}
		\caption{\label{Fig1} (a) Energy dispersion of the lowest two conduction bands of SrTiO$_3$ for a $kp$ model using the parameters obtained by Uwe {\it et al.} \cite{Uwe1985}, where $a$ is the length of the cubic SrTiO$_3$. The energy surfaces at $2.5$ meV for the (b) second-lowest and (c) lowest bands.}
	\end{center}
\end{figure}
%============================================================

Strontium titanate (SrTiO$_3$) is one such material that exhibits non-saturated MR with a single closed Fermi surface \cite{XLin2013,Allen2013}. SrTiO$_3$ crystalizes into a cubic perovskite structure at room temperature and undergoes a structural transition from a cubic structure to a tetragonal structure at $T=105$ K \cite{Collignon2018}. 
The band structure of the conduction band of tetragonal SrTiO$_3$ based on the $\kp$ model proposed by Uwe {\it et al.} \cite{Uwe1985} and Allen {\it et al.} \cite{Allen2013} is presented in Fig. \ref{Fig1} (a). 
Experimentally, the critical carrier density at which a second Fermi surface appears is approximately $n_c \simeq 1.2 \times 10^{18}$ cm$^{-3}$ \cite{XLin2014}. 
However, significant non-saturated linear transverse MR (over 300\% at 10 T) is experimentally observed below $n_c$, where there is only one closed Fermi surface \cite{XLin2013,Allen2013}.
The reason why such large transverse MR appears with only a single closed Fermi surface is still unknown, even though it is a very fundamental question. 

The purpose of this paper is to clarify the origin of the MR of SrTiO$_3$. Our main hypotheses are as follows. There are two possible origins of MR with a single closed Fermi surface: (i) a warped Fermi surface from a perfect sphere and/or (ii) thermally excited carriers. The common belief that a single closed Fermi surface does not generate MR is derived from models of perfect spheres and ellipsoids. However, for real metals, the Fermi surface is warped from a perfect sphere. In fact, the energy surface of SrTiO$_3$ deviates considerably from that of a perfect sphere, as shown in Fig. \ref{Fig1} (b) and (c). It is naively expected that this warping will contribute to MR. The problem is determining how large the MR will be and if it will be saturated. It is well known that semimetals with electron and hole carriers exhibit non-saturated MR when compensation is perfect. Similarly, a multi-valley system can also exhibit MR, but it will be saturated at strong fields \cite{ZZhu2018}. In the case of SrTiO$_3$, the thermally excited carriers in the second-lowest band should contribute to MR because the energy gap is very small ($\sim 1.5$ meV). However, the magnitude and saturation of the MR are unknown.

We first investigate these two questions separately, then merge them into a realistic model of SrTiO$_3$. MR is calculated based on the Boltzmann equation under a magnetic field. We extend the formula for MR proposed by Mackey and Sybert \cite{Mackey1969}, where only a simple ellipsoidal Fermi surface was studied, to calculate MR for an arbitrarily shaped Fermi surface. 
As a general model for warped Fermi surfaces, we calculate the MR for a Fermi surface expanded in terms of cubic harmonics. It is shown that the angular dependence of the diagonal elements in the mass tensor causes transverse MR, while that of the off-diagonal elements causes longitudinal MR, even when a Fermi surface is single and closed. It is also shown that thermally excited carries can contribute to the MR. This contribution is roughly proportional to temperature. Finally, the warped Fermi surface and the thermally excited carriers are considered simultaneously by calculating MR using a $k.p$ model of SrTiO$_3$. We conclude that the large MR of SrTiO$_3$ is mainly due to warping of the Fermi surface, but it is also affected by thermally excited carriers.
The effects of the negative Gaussian curvature of the Fermi surface and possible negative MR are also discussed.

\section{Extended Mackey-Sybert formula for MR}
We calculate MR using the formula derived by Mackey and Sybert \cite{Mackey1969}. Their formula is based on the Boltzmann equation under electric and magnetic fields. In the Mackey--Sybert formula, a magnetic field is expressed in matrix form as
\begin{align}
	\hat{B}=\left(
	\begin{array}{ccc}
		0 & -B_z & B_y \\
		B_z & 0 & -B_x \\
		-B_y & B_x & 0
	\end{array}
	\right).
\end{align}
The magnetic field tensor $\hat{B}$ corresponds to the standard electromagnetic tensor (i.e., the tensor itself is not novel). However, the formula for MR can be drastically simplified by using $\hat{B}$. This simplified formula enables us to unravel the mechanisms of MR with a single closed Fermi surface, as shown later in this paper.  
Unfortunately, this original formula is not applicable to a system with a complex Fermi surface, such as SrTiO$_3$ (Fig. \ref{Fig1}), because Mackey and Sybert only studied a case with an ellipsoidal Fermi surface.
In this paper, we extend the Mackey--Sybert formula to consider arbitrarily shaped Fermi surfaces. This new formulation is relatively straightforward. We simply maintain the wave number dependence of the energy dispersion throughout the calculation. The resulting formula for an MR tensor $\hat{\rho}$ is written as
\begin{align}
	\hat{\rho}&=\hat{\sigma}^{-1}
	\\
	\sigma_{\mu \nu}&=e\left\langle v_{k, \mu} \,
	\left\{ \bv_k \cdot \left[\left( \frac{1}{e\tau}-  \hat{B}\cdot \hat{\alpha}_k \right)^{-1}  \right] \right\}_\nu \right\rangle_F .
	\label{Conductivity}
\end{align}
Here, $\langle \cdots \rangle_F =\int(d\bk/4\pi^3)(\cdots)(-\partial f_0/\partial \epsilon)$ expresses integration along the Fermi surface at low temperatures ($f_0$ is the equilibrium Fermi distribution function). The velocity $\bv_k$ and inverse mass tensor $\hat{\alpha}_k$ are defined in terms of the energy dispersion $\ve_k$ as $v_{k, \mu} =\partial \ve_k / \hbar \partial k_\mu$ and $\alpha_{k, \mu \nu}=\partial^2 \epsilon_k /\hbar^2 \partial k_\mu \partial k_\nu$. These terms are independent of $\bk$ in the original Mackey--Sybert formula. In Eq. \eqref{Conductivity}, the relaxation time $\tau$ is expressed as a constant scalar, but can be expressed as a tensor that is dependent on momentum. This point will be discussed later.

\section{MR due to warp in a Fermi surface}
%============================================================
\begin{figure}
	\begin{center}
		\includegraphics[width=7cm]{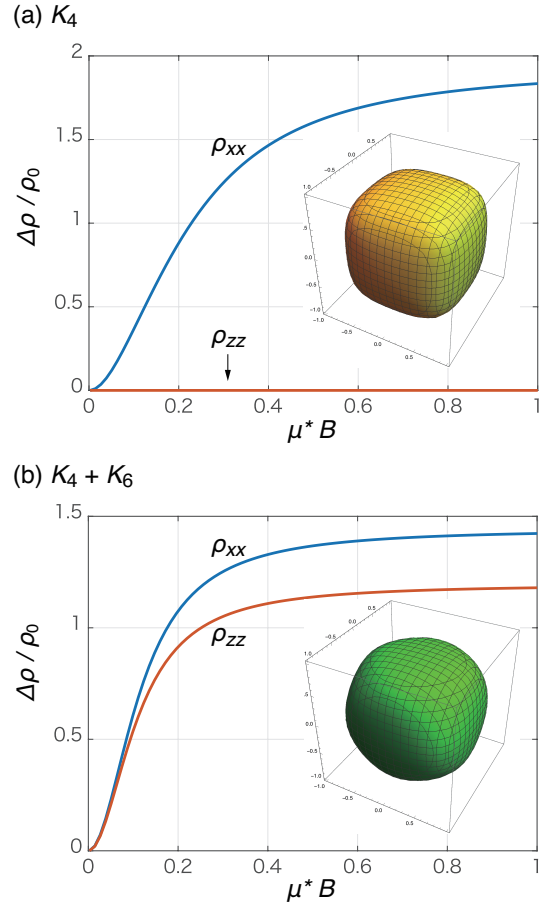}
		\caption{\label{Fig3} MR for the cubic harmonics model [Eq. \eqref{CHmodel}] with (a) $K_4$-term only: $(c_4, c_6)=(1.0, 0.0)$ and (b) $K_4$ and $K_6$ terms: $(c_4, c_6)=(1.0, 0.1)$. The insets show their Fermi surfaces. The magnetic field $B$ is scaled by the effective mobility $\mu^*=e\tau/m^*$.}
	\end{center}
\end{figure}
%============================================================

First, we examine the effects of warp in Fermi surface. We assume that energy is described in terms of cubic harmonics as \cite{Olson1957}
\begin{align}
	\ve_k = \frac{\hbar^2 k_0^2}{2m^*}
	\Biggl[
	\tilde{k}^2
	+c_4 \tilde{k}^4 K_4 (\theta, \varphi) 
	+ c_6 \tilde{k}^6 K_6 (\theta, \varphi)
	\Biggr],
	\label{CHmodel}
\end{align}
where $\tilde{k}^2=k^2/k_0^2
$ and $K_n (\theta, \varphi)$ defines cubic harmonics of degree $n$, which are written as
\begin{align}
	K_4 (\theta, \varphi) &= \frac{5}{2}\left( \xi^4 + \eta^4 + \zeta^4 -\frac{3}{5}\right),
	\\
	K_6 (\theta, \varphi) &=
	\frac{231}{2}\left[
	\xi^2 \eta^2 \zeta^2 + \frac{1}{55}K_4 (\theta, \varphi)-\frac{1}{105}
	\right].
\end{align}
($\xi=\sin \theta \cos \varphi$, $\eta=\sin \theta \sin \varphi$, $\zeta = \cos \theta$) The obtained Fermi surfaces are depicted in the insets in Fig. \ref{Fig3}. The Fermi surface with only the $K_4$ term is largely cubic, whereas the $K_6$ term adds a cuboctahedron component.

In Fig. \ref{Fig3}, the calculated transverse ($\Delta \rho_{xx}$) and longitudinal ($\Delta \rho_{zz}$) MR for the cubic harmonics model are plotted as a function of dimensionless $\mu^* B$, where the magnetic field is oriented along the $z$-axis, $\bm{B}=(0, 0, B)$, and $\mu^*=e\tau/m^*$ is the effective mobility. The results with only $K_4$ are presented in Fig. \ref{Fig3} (a) and those with both $K_4$ and $K_6$ are presented in Fig. \ref{Fig3} (b). A remarkable $\Delta \rho_{xx}$ value is obtained for both cases. In the weak field region, $\Delta \rho_{xx} \propto B^2$. Additionally, it is saturated in the strong field. In the intermediate region, it behaves as if $\Delta \rho_{xx} \propto B$. The magnitude of $\Delta \rho_{xx} /\rho_0$ becomes larger in proportion to $c_4$ and $c_6$, where $\Delta \rho_{\mu\mu}=\rho_{\mu\mu}-\rho_0$ and $\rho_0$ is the resistivity at zero field.
Furthermore, $\rho_{zz}$ becomes finite when the Fermi surface exhibits cubic harmonics higher than $n=6$. No longitudinal MR appears with only the $K_4$ warp.

The above properties of the cubic harmonics model can be interpreted as follows.
In the case with the $K_4$ term, $\hat{\alpha}_k$ is diagonal, which is the same situation as the ellipsoidal model. Regardless, the $K_4$ warp generates $\rho_{xx}$, while the ellipsoidal model does not. The difference between these models lies in the angular dependence of $\hat{\alpha}_k$.
The magnetoconductivities are written as
\begin{align}
	\sigma_{xx}&=e\left\langle\frac{v_x^2/e\tau}{(e\tau)^{-2}+\alpha_{xx}\alpha_{yy}B^2} \right\rangle_F ,
	\\
	\sigma_{xy}&=e\left\langle\frac{-v_x^2\alpha_{yy}B}{(e\tau)^{-2}+\alpha_{xx}\alpha_{yy}B^2} \right\rangle_F 
\end{align}
for $\bm{B}=(0, 0, B)$.
($\sigma_{yy}$ and $\sigma_{yx}$ are obtained through the exchange of $(x\leftrightarrow y)$ in $\sigma_{xx}$ and $\sigma_{xy}$, respectively.)
The transverse MR is written as
\begin{align}
	\rho_{xx}=\frac{\sigma_{yy}}{\sigma_{xx}\sigma_{yy}-\sigma_{xy}\sigma_{yx}}.
	\label{rhoxx}
\end{align}
In the case where $\hat{\alpha}$ and $\bm{v}$ are constant along the Fermi surface (ellipsoidal model), we can drop $\langle \cdots \rangle_F$. Then, the field dependence of the numerator in $\rho_{xx}$ is cancelled out by that of the denominator, resulting in a lack of field dependence in the transverse direction. In contrast, in the case with the $K_4$ term, such a cancellation does not occur because we cannot drop $\langle \cdots \rangle_F$ based on the angular dependence of $\hat{\alpha}_k$. This is why the $K_4$ term generates $\Delta \rho_{xx}$. However, in the case with only the $K_4$ term, $\Delta \rho_{zz}$ is independent of $B$ because
\begin{align}
	\sigma_{zz}=e^2\left\langle \tau k_z^2 \right\rangle_F.
\end{align}
Therefore, $\Delta \rho_{zz}$ cannot be generated with only the $K_4$ term.

The $K_6$ term gives the off-diagonal elements in $\hat{\alpha}_k$, meaning $\sigma_{zz}$ depends on $B$ as
\begin{align}
	\sigma_{zz}&=e^2 \tau\left\langle S_1/S_2
	\right\rangle_F
	\label{szz}
	\\
	S_1&=
	v_z^2 (e\tau)^{-2}
	+\bigl[v_z^2 (\alpha_{xx} \alpha_{yy} -\alpha_{xy}^2)
	\nonumber\\&
	\hspace{18mm}+ v_z v_x (\alpha_{xy} \alpha_{yz} -\alpha_{yy} \alpha_{zx})
	\nonumber\\&
	\hspace{18mm}+ v_y v_z (\alpha_{xy} \alpha_{zx}-\alpha_{xx} \alpha_{yz}) 
	\bigr]B^2
	\label{S1}
	\\
	S_2&=
	\left[ (e\tau)^{-2} + (\alpha_{xx}\alpha_{yy}-\alpha_{xy}^2)B^2\right].
	\label{S2}
\end{align}
When $\hat{\alpha}$ is constant along the Fermi surface, the terms that include $v_z v_x$ or $v_y v_z$ will disappear during integration with respect to $\bk$ because they are odd in $k_\mu$. The field dependencies are canceled between the numerator and denominator, resulting in $\sigma_{zz}=e^2\langle \tau k_z^2\rangle_F$. However, when $\hat{\alpha}_k$ has an angular dependence, as in the case with $K_6$, the $v_z v_x$ and $v_y v_z$ terms in Eq. \eqref{S1} provide finite contributions (each term is even with respect to $k_\mu$), meaning the field dependence remains. In other words, the factors $(\alpha_{xy} \alpha_{yz} -\alpha_{yy} \alpha_{zx})$ and $(\alpha_{xy} \alpha_{zx}-\alpha_{xx} \alpha_{yz})$ are the source of $\Delta \rho_{zz}$. This is why longitudinal MR appears only with $K_6$ term, and does not appear with $K_4$ alone.

\section{MR due to thermally excited carriers}
%============================================================
\begin{figure}
	\begin{center}
		\includegraphics[width=7cm]{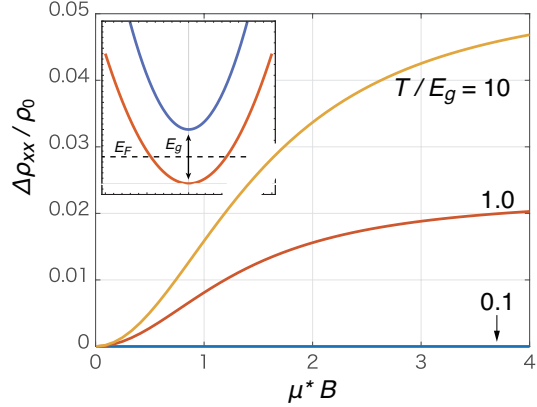}
		\caption{\label{Fig4} Transverse MR of the free-electron two-band model for $T/E_g=0.1, 1.0, 10$. The inset shows the band structure of the model, where the Fermi energy is set to $E_F=E_g/2$.}
	\end{center}
\end{figure}
%============================================================
We now examine the effects of thermally excited carriers. We consider a model where two free-electron bands are separated by an energy $E_g$. The ratio of the effective mass for the upper band to that for the lower band is 0.6, which was set to be consistent with SrTiO$_3$. The Fermi energy $E_F$ lies between the band edges, as shown in the inset in Fig. \ref{Fig4}. The $\Delta \rho_{xx}$ for this model are presented in Fig. \ref{Fig4}. Again, $\Delta \rho_{xx}\propto B^2$ at weak fields, whereas it is saturated at strong fields. $\Delta \rho_{xx} \propto B$ in the intermediate region. 

This $\Delta \rho_{xx}$ is due to the thermally excited carries, which make the system a quasi-two-carrier system, even though there is only a single closed Fermi surface. It is known that multiple carrier systems, such as semimetals or multivalley systems, show $\Delta \rho_{xx}$, but do not show any $\Delta \rho_{zz}$ \cite{ZZhu2018}. Therefore, only $\Delta \rho_{xx}$ appears in the considered two-band model. At sufficiently low temperatures, $T/E_g \ll 1$, $\Delta \rho_{xx}$ is suppressed because the thermally excited carriers virtually disappear.

\section{MR of SrTiO$_3$}
\subsection{$k.p$ model for SrTiO$_3$}
We now calculate the MR of SrTiO$_3$. We employ the $k.p$ model proposed by Uwe {\it et al.} \cite{Uwe1985}. This model was derived more rigorously by Allen {\it et al.} based on the model proposed by Khalsa and MacDonald \cite{Khalsa2012,Allen2013}. Its energy dispersion is written as
\begin{align}
	E_\pm &=\frac{\gamma_1}{2}k^2 \pm 
	\bigl[\gamma_2^2 k^4 - 3(\gamma_2^2-\gamma_3^2)(k_x^2 k_y^2 + k_y^2 k_z^2 + k_z^2 k_x^2)
	\nonumber\\&
	+\gamma_2 be(2k_z^2 -k_x^2 -k_y^2) + (b e)^2
	\bigr]^{1/2} +|be|,
	\label{kp}
\end{align}
where $k_\mu$ is normalized by $\pi/a$ ($a$ is the lattice constant along the $a$-axis).
$E_-$ and $E_+$ correspond to the energy of the lowest and second-lowest bands, respectively. The origin of the energy is considered to be the bottom of $E_- $.
This $k.p$ model considers the spin-orbit splitting and tetragonal distortion at low temperatures caused by the tetragonal strain $e$ and deformation potential $b$, respectively. 
Uwe {\it et al.} set the parameters for this equation to fit their experimental results for angle-resolved Shubnikov-de Haas (SdH) oscillations. Allen {\it et al.} confirmed the results of Uwe {\it et al.} by the SdH measurement with thin films \cite{Allen2013}. In the following discussion, we show the results of the model with the parameters proposed by Uwe {\it et al.}: $\gamma_1=3.5$ eV, $\gamma_2$=0.88 eV, $\gamma_3=0.13$ eV, and $be=-0.75$ meV (the results are nearly the same when using the parameters proposed by Allen {\it et al.}). 
The resulting energy surfaces are presented in Fig. \ref{Fig1}. They clearly deviate from those of a perfect sphere. 

%============================================================
\begin{figure}
	\begin{center}
		\includegraphics[width=7cm]{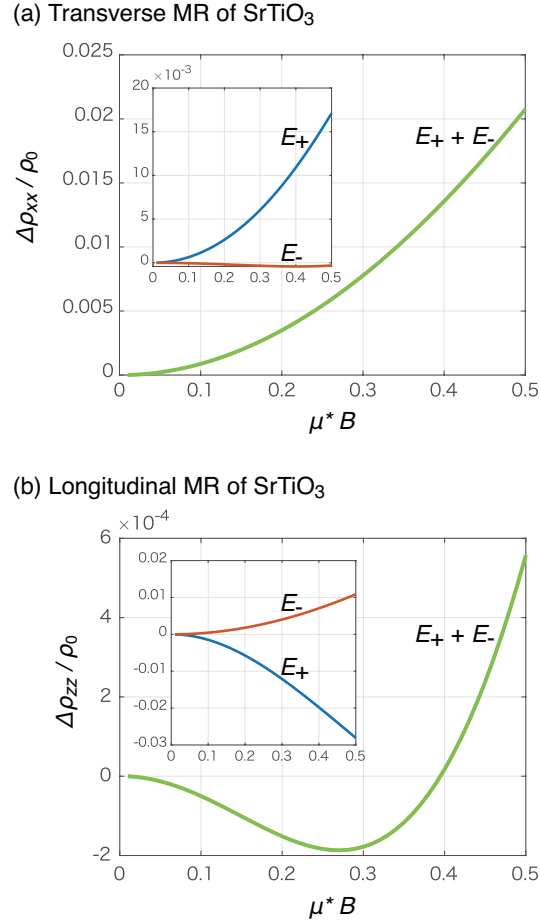}
		\caption{\label{Fig7} (a) Transverse and (b) longitudinal MR for SrTiO$_3$ model with the constant relaxation time at $T=15$ K. ($E_F=1.3$ meV and $\tau=3.0 \times 10^{-12}$ s.) The insets shows the MR for individual bands $E_\pm$.}
	\end{center}
\end{figure}
%============================================================

The calculated MR values for this SrTiO$_3$ model are presented in Fig. \ref{Fig7}. The Fermi energy was set to $E_F=1.3$ meV, which gives the carrier density $n=3.9\times 10^{-17}$ cm$^{-3}$. ($n$ is calculated from the Hall coefficient.) Clear MR signals were obtained in both the transverse and longitudinal directions. This can be easily explained according to the insights obtained from the cubic harmonics model.
The $k^4$ term in Eq. \eqref{kp} yields the angle-dependent diagonal elements in $\hat{\alpha}_k$, whereas the $k_\mu^2 k_\nu^2$ term yields the angle-dependent off-diagonal elements. The former causes $\Delta \rho_{xx}$ and the latter causes $\Delta \rho_{zz}$. 
The MR of each band is presented in the inset in Fig. \ref{Fig7}. By comparing the results of the composite band to those of the individual bands, one can see that the total MR is largely modified from the MR of $E_-$. The thermally excited carriers in $E_+$ lead to an increase in $\Delta \rho_{xx}^{\rm total}$ and decrease in $\Delta \rho_{zz}^{\rm total}$. The contribution of the thermally excited carriers is reduced by decreasing temperature, i.e., $\Delta \rho^{\rm total}$ approach $\Delta \rho$ of $E_-$.

It is revealed that $\Delta \rho_{xx}$ of $E_-$ and $\Delta \rho_{zz}$ of $E_+$ become negative, which cannot be explained based on the simple understanding of the cubic harmonics model. However, the negative MR values can be explained if one considers the fine structures of the Fermi surface, such as the Gaussian curvature.

\subsection{Gaussian curvature}
In the conductivity expression, a factor such as $S_2$ appears in the denominator, as shown in Eq. \eqref{szz}. The coefficient $(\alpha_{xx}\alpha_{yy}-\alpha_{xy}^2)$ exactly coincides with the Gaussian curvature of the energy surface (except for the normalization factor). The Gaussian curvature is given as the product of two principal curvatures. Intuitively, the Gaussian curvature becomes negative when the curvature along one direction is positive and that along the other direction is negative. For example, the saddle point has negative Gaussian curvature.
The Gaussian curvature appears in various physical quantities, e.g., the Landau-Peierls formula of diamagnetism \cite{Peierls1933b,Peierls_book} or spin Hall effect \cite{Fuseya2015}. 

\begin{figure}
	\begin{center}
		\includegraphics[width=7cm]{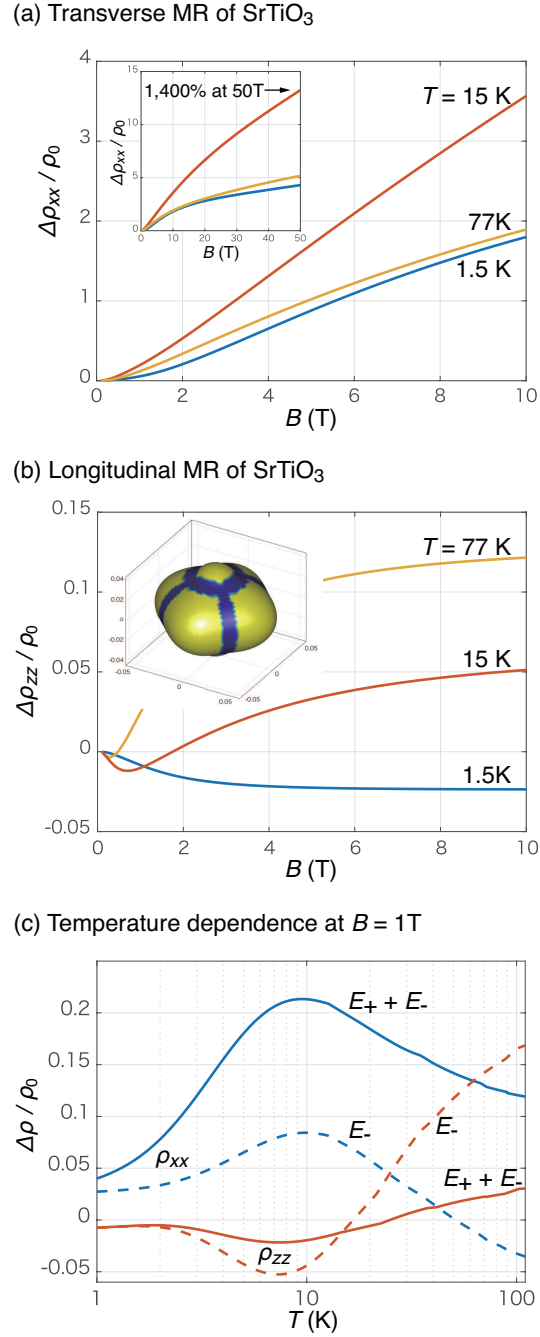}
		\caption{\label{Fig5} (a) Transverse and (b) longitudinal MR for the SrTiO$_3$ model ($E_F=1.3$ meV, i.e., $n=3.9\times 10^{-17}$ cm$^{-3}$) with the anisotropic relaxation time ($\tau_0=0.9\times 10^{-11}$ s, and $\alpha_0 =0.91/ m_e$ for $E_+$ and $\alpha_0 = 0.54/m_e$ for $E_-$.) at $T=1.5, 15, 77$ K. Note that $\tau_0$ is assume to be constant, although it depends on temperature in general. The inset in (a) shows the transverse MR at strong fields. The increase of $\rho_{xx}$ (15 K) reaches 450 \% at 30 T and 1,400 \% at 50 T. The inset in (b) highlights the region where the Gaussian curvature is negative for the energy surface of $E_-$. 
		(c) Temperature dependence of the transverse and longitudinal MR. The solid lines are the results for the combined band, $E_+ +E_-$, while the dashed lines are those for the lowest band, $E_-$.}
	\end{center}
\end{figure}

When the Gaussian curvature is negative, the effects of the magnetic field are reversed. Specifically, the magnetoconductivity is increased by the field, resulting in negative MR.
In the case of SrTiO$_3$, the dimples in the energy surface are a source of negative Gaussian curvature. In the inset in Fig. \ref{Fig5} (b), the region where the Gaussian curvature is negative is highlighted. Based on this negative Gaussian curvature, $\Delta \rho_{xx}$ of $E_-$ exhibits negative MR, as shown in the inset in Fig. \ref{Fig7}. The sign of $\Delta \rho_{zz}$ is more complex. It is also affected by the signs in the denominator, specifically the signs of $(\alpha_{xy}\alpha_{yz}-\alpha_{yy}\alpha_{zx})$ and $(\alpha_{xy}\alpha_{zx}-\alpha_{xx}\alpha_{yz})$. 

In a case with negative Gaussian curvature, $S_2$ can become zero at a certain strength of field, meaning the magnetoconductivities can diverge. However, this divergence is an artificial phenomenon. For example, divergence can be due to the approximation of the constant $\tau$ in Eq. \eqref{Conductivity}. Divergence does not occur when we introduce an anisotropic $\tau$. Many authors have highlighted the importance of the anisotropy of $\tau$ by introducing a relaxation time tensor $\hat{\tau}$ \cite{Herring1956,Hubner1967,Hartman1969,Mackey1969,Fuchser1970,Collaudin2015,ZZhu2018}. One of the simplest and most naive definitions would be $\hat{\tau}=(\tau_0/\alpha_0)\hat{\alpha}$, where $\tau_0$ and $\alpha_0$ are the average relaxation time and average inverse mass, respectively. This relationship is easily obtained from the assumption that the mean free path is comparable to the Fermi wavelength $\lambda_F \propto \alpha_0$.

%============================================================
%============================================================

Figure \ref{Fig5} presents the MR results for SrTiO$_3$ after introducing $\hat{\tau}=(\tau_0/\alpha_0)\hat{\alpha}$ into Eq. \eqref{Conductivity} with $\tau_0=0.9\times 10^{-11}$ s, and $\alpha_0 =0.91/ m_e$ for $E_+$ and $\alpha_0 = 0.54/m_e$ for $E_-$. ($m_e$ is the bare electron mass.) These parameters were obtained to match the zero field values in previous experiments. A remarkable $\Delta \rho_{xx}$ was obtained. The increase in $\Delta \rho_{xx}$ is approximately 300\% at 10 T and 1.5 K, which is comparable to the values obtained experimentally: $\sim 300\%$ from Allen {\it et al.} at 0.41 K \cite{Allen2013} and $\sim 470\%$ from Lin {\it et al.} at 1.2 K \cite{XLin2013}. (Note that $\Delta \rho/\rho_0 =2$ corresponds to an increase of 300 \%) Furthermore, one can see that $\Delta \rho_{xx} \propto B$ up to 10 T, which is consistent with the experimentally observed linear MR for SrTiO$_3$. According to our calculations, the MR keeps the quasi-linear field-dependence up to 50 T (MR $\sim 1,400\%$) at 15 K as shown in the inset of Fig. \ref{Fig5}. The apparent linear MR is a result of the applied field being not to reach the saturation field. %Note that the data up to 30 T presented by Allen {\it et al.} exhibit a tendency to become saturated.

We observed negative longitudinal MR as is shown in Fig. \ref{Fig5} (b), but these results have not been experimentally confirmed for SrTiO$_3$. As discussed above, the field dependence of $\Delta \rho_{zz}$ originates from the terms $(\alpha_{xy}\alpha_{yz}-\alpha_{yy}\alpha_{zx})$ and $(\alpha_{xy}\alpha_{zx}-\alpha_{xx}\alpha_{yz})$ in Eq. \eqref{S1}. When these terms make positive contributions, $\sigma_{zz}$ is an increasing function with respect to $B$, resulting in a negative $\Delta \rho_{zz}$. Recently, negative longitudinal MR has been discussed in connection with the topology of materials and the chiral anomaly \cite{Burkov2015,Reis2016,Arnold2016,JXu2019}. The negative longitudinal MR discussed here is not related to topology or chiral anomaly, but is due to the warp in Fermi surfaces. 

The temperature dependences of $\Delta \rho_{xx}$ and $\Delta \rho_{zz}$ at $B=1$ T are shown in Fig. \ref{Fig5} (c). The solid lines are the results for the combined band $E_+ + E_-$, and the dashed lines are those for the lowest band $E_-$. Both $\Delta \rho_{xx}$ and $\Delta \rho_{zz}$ are peaked at around 10 K. The negative contribution above 10 K originates from the negative Gaussian curvature around $E\sim2.3$ meV in $E_-$ [cf. Fig. \ref{Fig1} (a)]. (The excitation energy to that energy region is about 1 meV since $E_F=1.3$ meV.) The contributions from the thermally excited carriers can be evaluated by subtracting $\Delta \rho$ for $E_-$ from that for $E_+ + E_-$, i.e., the distance between the solid and dashed lines. It is clear from Fig. \ref{Fig5} (c) that the thermally excited carriers boost the MR. The enhancement  in $\Delta \rho_{xx}$ by the thermally excited carrier is 190\% at 1.5 K and 260\% at 15 K. For $\Delta \rho_{zz}$, the contribution from the thermally excited carriers essentially increases as increasing temperature, although the temperature dependence is more complex due to the opposite-sign contributions between $E_+$ and $E_-$. Here are some points to be noted. The relaxation time $\tau_0$ in this calculation is assumed to be independent from temperature, though it depends on temperature as $\tau_0 \propto T^{-2}$ in general \cite{Hartman1969,Collaudin2015,Fauque2018}. If one considers this temperature dependence, $\Delta \rho$ monotonously increases as decreasing temperature.

Lastly, it would be worth mentioning the effect of the Zeeman splitting, which can increase the amplitude of MR as has been studied in various systems \cite{GXiong2000,Peters2010,Kajita2014}. The total carrier density at $E_F$ will decrease when the band energy for the spin up and down are split by the Zeeman effect. Then the resistivity will increase since it is inversely proportional to the carrier density. The effect becomes the largest when the Zeeman energy exceeds $E_F$. In the case of SrTiO$_3$, this condition will be satisfied around $B \sim 40$ T for $E_-$. Therefore, for $B \lesssim 10$ T, the increase of MR is expected to be not so large (less than a few percent\cite{GXiong2000}). For $B \gtrsim 10$ T, on the other hand, the Zeeman split band of $E_+$ will increase the carrier density at $E_F$ and contribute to the MR negatively. As a result, the increase of MR due to the Zeeman splitting will be reduced.

\section{Conclusions}
We studied both the transverse $\Delta \rho_{xx}$ and longitudinal $\Delta \rho_{zz}$ MR of SrTiO$_3$, which has a single closed Fermi surface. We extended the Mackey--Sybert formula for MR to make it applicable to arbitrarily shaped energy surfaces. We examined two possible origins for MR: (i) warp in Fermi surfaces and (ii) thermally excited carriers. It was shown that the angular dependence of the diagonal components of the inverse mass tensor $\hat{\alpha}_k$ causes $\Delta \rho_{xx}$, but not $\Delta \rho_{zz}$. In contrast, the angular dependence of the off-diagonal components of $\hat{\alpha}_k$ causes $\Delta \rho_{zz}$. Although thermally excited carries enhance the magnitude of the MR, their contribution is reduced by decreasing temperature.

We calculated MR using a $k.p$ model for SrTiO$_3$ proposed by Uwe {\it et al.} and Allen {\it et al.} Clear $\Delta \rho_{xx}$ and $\Delta \rho_{zz}$ were obtained, even though there is only one closed Fermi surface for this material. The increase in $\rho_{xx}$ in our calculations was approximately 300 \% at 10 T, which quantitatively agrees with experimental values.  $\Delta \rho_{xx}$ exhibits the apparent linear field-dependence even at 50 T (MR $\sim 1,400 \%$). The potential for negative MR as a result of negative Gaussian curvature was discussed. Our calculations indicate that negative longitudinal MR is possible for SrTiO$_3$. This observation will be considered in future research.

\section*{Acknowledgments}
We would like to thank K. Behnia, B. Fauqu\'e, and C. Collignon for their helpful insight and for providing experimental data prior to its publication. We also thank Y. Yanase for his helpful comments.
This work is supported by JSPS KAKENHI grants 16K05437 and 15KK0155.

\section*{References}
\bibliographystyle{iopart-num}
\bibliography{SrTiO3}

\end{document}